\documentclass[12pt]{article}
\usepackage{epsfig}
\usepackage{amsmath}
\usepackage{hhline}
\usepackage{amssymb}
\usepackage{times}
\usepackage{rotating}

\newlength{\dinwidth}
\newlength{\dinmargin}
\newcommand{\GeV}{\rm GeV}

\def\gsim{\,\lower.25ex\hbox{$\scriptstyle\sim$}\kern-1.30ex%
\raise 0.55ex\hbox{$\scriptstyle >$}\,}
\def\lsim{\,\lower.25ex\hbox{$\scriptstyle\sim$}\kern-1.30ex%
\raise 0.55ex\hbox{$\scriptstyle <$}\,}

\newcommand{\gev}{\,\mbox{GeV}}

\def\EJC{{\em Eur. Phys. J.} {\bf C}}

\newcommand{\thstarhad}{${\theta_W^{\; \bar{q}}}$~}
\newcommand{\thstarlep}{${\theta_W^{\; \ell}}$~}
\newcommand{\costarhad}{$\cos{\theta_W^{\; \bar{q}}}$~}
\newcommand{\costarlep}{$\cos{\theta_W^{\; \ell}}$~}

\newcommand{\ptb}{$P_T^b$~}
\newcommand{\mtopl}{$M_{\ell \nu b}$~}

\begin{document}

\pagestyle{empty}
\begin{titlepage}

\noindent
DESY-03-132 \hspace{9cm} ISSN 0418-9833\\
October 2003  \\
\vspace*{3cm}

\begin{center}
  \Large {\boldmath \bf  Search for Single Top Quark Production \\ in $ep$ Collisions at HERA}

  \vspace*{1cm}
    {\Large H1 Collaboration} 
\end{center}

\begin{abstract}
\noindent
%%%%%%%%%%%%%%%%%%% ABSTRACT%%%%%%%%%%%%%%%%%%%%%
 A search for single top quark production is performed in $e^{\pm}p$ collisions at HERA.
 The search exploits data corresponding to an integrated luminosity of 118.3~pb$^{-1}$.
A model for the anomalous production of top quarks in a flavour changing neutral current process involving a $tu\gamma$ coupling is investigated.
 Decays of top quarks into a \mbox{$b$ quark} and a $W$ boson are considered in the leptonic and the hadronic decay channels of the~$W$.
Both a  cut--based analysis and a multivariate likelihood analysis are performed to discriminate anomalous top quark production from Standard Model background processes.
 In the leptonic channel, 5~events are found
while 1.31 $\pm$ 0.22 events are expected from the Standard Model background.
 In the hadronic channel, no excess above the expectation for Standard Model processes is found. 
%Due to the small number of top candidates, upper limits of 0.55~pb on the anomalous top production cross section at $\sqrt{s}=319$~\gev~and 0.27 on the $tu\gamma$ coupling $\kappa_{tu\gamma}$ are established at the 95\%~confidence level.
These observations lead to a cross section $\sigma (ep \rightarrow e \, t X)
= 0.29 \, ^{+0.15}_{-0.14} \ {\rm pb}$ at $\sqrt{s} = 319 \ {\rm GeV}$. 
Alternatively, assuming that the observed events are due to a statistical
fluctuation, upper limits of 0.55~pb on the anomalous top production cross section and of 0.27 on the $tu\gamma$ coupling $\kappa_{tu\gamma}$ are established at the 95\%~confidence level.

\end{abstract}

\vfill
\begin{center}
Submitted to \EJC
\end{center}

\end{titlepage}

\begin{flushleft}
  %-- H1AUTS Author list by names 
%-- Status: Thu Feb  6 11:18:53 MET 2003  Number of authors = 299 

A.~Aktas$^{10}$,               %DESY-ST        03/2            Aktas               
V.~Andreev$^{24}$,             %LPI -PD        8/88            Andreev             
T.~Anthonis$^{4}$,             %ANTW-ST        11/99           Anthonis            
A.~Asmone$^{31}$,              %ROME-ST        07/2            Asmone              
A.~Babaev$^{23}$,              %ITEP-PD        8/88            Babaev              
S.~Backovic$^{35}$,            %ZEUT-PD        03/2            Backovic            
J.~B\"ahr$^{35}$,              %ZEUT-PD        8/88            Baehr               
P.~Baranov$^{24}$,             %LPI -PD        8/88            Baranovp            
E.~Barrelet$^{28}$,            %PARI-PD        11/99           Barrelet            
W.~Bartel$^{10}$
% ...

%-- H1 Institutes 
\bigskip{\it
 $ ^{1}$ I. Physikalisches Institut der RWTH, Aachen, Germany$^{ a}$ \\
% ...
$ ^{4}$ Inter-University Institute for High Energies ULB-VUB, Brussels;
          Universiteit Antwerpen (UIA), Antwerpen; Belgium$^{ c}$ \\
% ...
$ ^{10}$ DESY, Hamburg, Germany \\
% ...
$ ^{23}$ Institute for Theoretical and Experimental Physics,
          Moscow, Russia$^{ l}$ \\
 $ ^{24}$ Lebedev Physical Institute, Moscow, Russia$^{ e}$ \\
% ...
$ ^{28}$ LPNHE, Universit\'{e}s Paris VI and VII, IN2P3-CNRS,
          Paris, France \\
% ...
$ ^{31}$ Dipartimento di Fisica Universit\`a di Roma Tre
          and INFN Roma~3, Roma, Italy \\
% ...
$ ^{35}$ DESY, Zeuthen, Germany \\
% ...
$ ^a$ Supported by the Bundesministerium f\"ur Bildung und Forschung, FRG,
      under contract numbers 05 H1 1GUA /1, 05 H1 1PAA /1, 05 H1 1PAB /9,
      05 H1 1PEA /6, 05 H1 1VHA /7 and 05 H1 1VHB /5 \\
 $ ^b$ Supported by the UK Particle Physics and Astronomy Research
      Council, and formerly by the UK Science and Engineering Research
      Council \\
 $ ^c$ Supported by FNRS-FWO-Vlaanderen, IISN-IIKW and IWT \\
% ...
$ ^e$ Supported by the Deutsche Forschungsgemeinschaft \\
% ...
$ ^l$ Partially Supported by Russian Foundation
      for Basic Research, grant    no. 00-15-96584 \\
\smallskip
% ...
}
\end{flushleft}
\newpage

\pagestyle{plain}

%%%%%%%%%%%%%%%%%%%%%%%%%%%%%%%%%%%%%%%%%%%%%%%%%%%%%%%%%%%%
%%%%% body
\section{Introduction}

In $ep$ collisions at the HERA collider, the production of single top quarks is kinematically possible due to the large centre--of--mass energy, which is well above the top production threshold.  
In the Standard Model, the dominant process for single top production at HERA is
the charged current reaction $e^+p \rightarrow \bar{\nu} t \bar{b} X$ ($e^-p \rightarrow \nu \bar{t} b X$). This process has a tiny cross section of less than 1~fb~\cite{Stelzer:1998ni,Moretti:1997dz} and thus Standard Model top production is negligible.
However, in several extensions of the Standard Model, the top quark 
is predicted to undergo flavour changing neutral 
current (FCNC) interactions, which could lead to a sizeable 
top production cross section. FCNC interactions are present in models which contain an extended Higgs sector~\cite{Atwood:1995ud},
Supersymmetry~\cite{deDivitiis:1997sh},  dynamical  
breaking of the electroweak symmetry~\cite{Peccei:kr}
or an additional symmetry~\cite{Fritzsch:1999rd}.
An observation of top quarks at HERA would thus be a clear indication of physics beyond the Standard Model. 
\par
The H1 Collaboration has reported~\cite{Adloff:1998aw,Andreev:2003pm} the observation of events with energetic isolated electrons and muons together with missing
transverse momentum in the positron--proton data collected between 1994 and 2000. 
The dominant Standard Model source is the production of real $W$ bosons. However, some of these events have a hadronic final state  with large transverse momentum, which is atypical of $W$ production. 
These outstanding events may indicate a production mechanism involving processes beyond the Standard Model.
 One such mechanism is the production of top quarks which predominantly decay into a $b$ quark and a $W$ boson.
The lepton and the missing transverse momentum would then be associated with a leptonic decay of the $W$ boson ($W \rightarrow \ell \nu$), while the observed high $P_T$ hadronic final state would be produced by the fragmentation of the $b$ quark.
 \par
 In this paper we present a search for anomalous single top production using 
 the H1~detector. The analysis uses the data collected between 1994 and 2000 corresponding to an integrated luminosity of $118.3$ pb$^{-1}$. The search covers the leptonic decay channel ($W \rightarrow \ell\nu$) and the hadronic decay channel ($W \rightarrow q \bar{q}'$) of the $W$ boson that emerges from the top quark decay. 
%($t \rightarrow b\ell\nu$: "leptonic channel", $t \rightarrow bq\bar{q'}$: hadronic channel). 
Both a cut--based analysis and a multivariate likelihood analysis are performed to select top quark candidates. 
Another search for single top production at HERA was recently published by the ZEUS Collaboration~\cite{zeus_singletop}.

%============================
\section{Phenomenology of FCNC Top Production}
%============================
\label{theory}

\begin{figure}[htb]
\centering
\epsfig{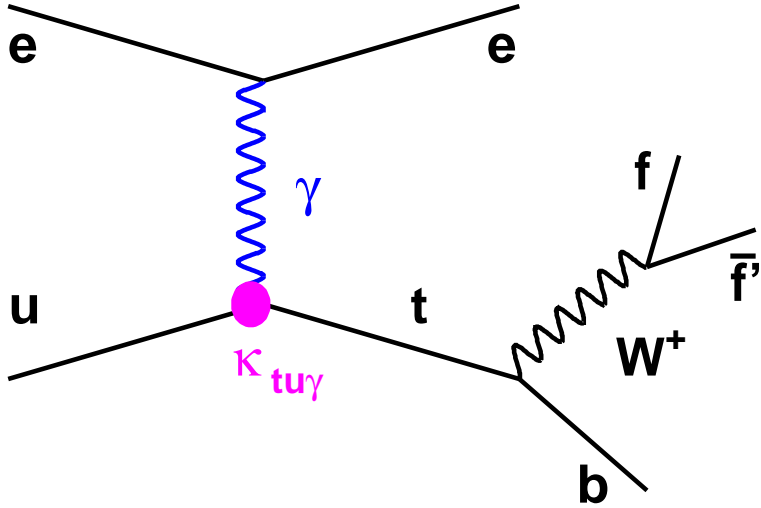} 
\caption{Anomalous single top production 
            via a flavour changing neutral current interaction  
            at HERA, with subsequent decays $t \rightarrow bW^+$ and  $W^+ \rightarrow f\bar{f'}$.
}
\label{fig:diagtop}
\end{figure}

A FCNC vertex involving the direct coupling of the top quark to a light quark ($u$ or $c$) and a gauge boson would lead to single top production, as
illustrated in figure~\ref{fig:diagtop}.
The most general effective Lagrangian, proposed 
in~\cite{Han:1998yr}, which describes FCNC top quark interactions involving
electroweak bosons is:
\begin{eqnarray}
\label{eq:lagrangian}
 {\cal{L}}_{eff}^{FCNC} &=&  \sum_{U = u,c}
  \frac{e e_U}{2\Lambda}  \kappa_{tU\gamma} \bar{t} \sigma_{\mu \nu} A^{\mu\nu} U \nonumber \\
 &+& \frac{g}{2 \cos \theta_W} \bar{t}
 \left[ \gamma_{\mu} (v_{tUZ} - a_{tUZ} \gamma^5) U Z^{\mu} +
  \frac{1}{2\Lambda}  \kappa_{tUZ}\sigma_{\mu \nu} Z^{\mu\nu} U  \right] 
 \quad + {\mbox {h.c.}}  \ ,
\end{eqnarray}
where $\sigma_{\mu \nu} = (i/2) \left[ \gamma^{\mu}, \gamma^{\nu} \right]$,
$\theta_W$ is the Weinberg angle, 
$e$ and $g$ are the couplings to the gauge groups with $U(1)$ and
$SU(2)$ symmetries, respectively, 
$e_U$ is the electric charge of up--type quarks, 
$A^{\mu}$ and $Z^{\mu}$ are the fields of the photon and the $Z$ boson
and $\Lambda$ is the scale up
to which the effective theory is assumed to be valid.
The tensors $A^{\mu\nu}$ and $Z^{\mu\nu}$ are the field strength tensors of the photon and $Z$ boson fields.
By convention $\Lambda$ is set  equal to the top mass in the following.
Because gauge invariance is required, only magnetic
operators allow FCNC couplings of the top quark to a photon and an up--type quark $U=u,c$ , denoted by $\kappa_{tU\gamma}$, while 
the non--vanishing $Z$ mass allows both the magnetic coupling $\kappa_{tUZ}$ and the vector and axial vector couplings to a $Z$ boson and an up--type quark, denoted by $v_{tUZ}$ and $a_{tUZ}$. 
\par
In $ep$-collisions, due to the large $Z$ mass, the contribution
of the $Z$ boson and the $\gamma-Z$ interference are  highly suppressed. Single top production is thus dominated by the 
$t$--channel exchange of a photon.  Therefore $Z$--exchange is neglected in this analysis and
only the $\kappa_{tU\gamma}$ couplings are considered.
\par
In the Standard Model FCNC processes can only arise via 
higher order corrections and are strongly suppressed. 
The possibility of anomalous single top production at HERA was first 
investigated
in~\cite{Fritzsch:1999rd}, where a model was built in which 
$\kappa_{tq\gamma} \propto m^2_{q}$ and therefore only the $tc \gamma$ 
coupling was relevant. 
In order to cover a large range of possible underlying theories, the model described by~(\ref{eq:lagrangian})
allows both couplings $\kappa_{tc\gamma}$ and $\kappa_{tu\gamma}$ to be present and independent of each other.
The sensitivity of HERA is much larger to the coupling
$\kappa_{tu\gamma}$ than to $\kappa_{tc\gamma}$ since the
$u$--quark density in the proton is much larger than the \mbox{$c$--quark} density at the high Bjorken~$x$ values needed to produce top quarks.
In the analysis presented in this paper only the coupling $\kappa_{tu\gamma}$ to the $u$--quark is considered.
\par
Compared with the production of top quarks in a $\gamma$--$u$ fusion process, the  corresponding  charge conjugate  anti--top quark production in a $\gamma$--$\bar{u}$ fusion process involving sea quarks is suppressed by a factor of $\sim$80. It is not considered in this analysis. 
\par
The simulation of the anomalous single top signal relies on an
event generator (ANOTOP), which uses the leading order (LO) matrix elements of the complete
$e + q \rightarrow e + t \rightarrow e + b + W \rightarrow e + b + f + \bar{f'}$
process as obtained from the CompHEP~\cite{COMPHEP} program.
The BASES/SPRING~\cite{Kawabata:1985yt} package is used to perform the numerical integration of the amplitudes and to generate events according to the resulting differential cross section. 
The parton shower approach~\cite{JETSET74}, which relies
on the leading logarithmic DGLAP~\cite{DGLAP} evolution equations, is used
to simulate QCD corrections in the initial and final states.
The MRST LO parton densities are used for the proton~\cite{Martin:1998sq}. The parton densities are evaluated at the top mass scale.
The nominal top mass is set to 175~GeV. A variation of the top mass by $+5$~GeV ($-5$~GeV) induces a cross section variation of $-20\%$ ($+25\%$).
The cross section calculation for anomalous single top production has recently been improved by including next--to--leading order (NLO)  QCD corrections~\cite{Belyaev:2001hf}. The uncertainty related to the choice of the renormalisation and factorisation scales is reduced to about 5\%. These NLO QCD corrections increase the cross section by about $17\%$ and are taken into account as an overall correction factor to the LO calculation for the results derived in this analysis. 

%======================================
\section{Standard Model Background Processes}
%======================================

Signatures of single top production are searched for in the leptonic and hadronic decay channels of the $W$ boson that emerges from the top quark decay. The relevant final state topology for the leptonic channel is an isolated lepton with high transverse momentum, at least one jet and missing transverse momentum. For the hadronic channel, the signature is three or more jets with high transverse momenta. The Standard Model processes that produce events with similar topologies and thus constitute the background for the present analysis are outlined below. Due to its small cross section, Standard Model top production is not considered in this analysis.
\par
The Standard Model processes that produce background to the leptonic channel were investigated in detail in the context of the isolated electron and muon analyses published by the H1 Collaboration in~\cite{Andreev:2003pm} and are only briefly described here. The main contribution is the production of $W$ bosons with subsequent leptonic decay of the $W$. 
The production of the electroweak
 vector bosons $W^{\pm}$ is modelled using the EPVEC~\cite{EPVEC} generator.
The NLO  QCD corrections to $W$ production~\cite{SPIRA} are taken into account
 by weighting the events as a function of the rapidity and transverse momentum of the $W$ boson~\cite{PRIVSPI}.
Other processes can contribute to the investigated final state through misidentification of photons or hadrons as leptons or through fake missing transverse momentum due to measurement fluctuations. Processes with a genuine lepton but possible fake missing transverse momentum are lepton pair production and neutral current (NC) deep inelastic scattering. The contribution from lepton pair production, dominated by two--photon processes where 
one of the two produced leptons is not detected, is calculated with the 
GRAPE~\cite{GRAPE} generator. 
The background contribution from NC deep inelastic scattering is estimated using the RAPGAP~\cite{RAPGAP} generator.  
In charged current (CC) deep inelastic scattering, the missing transverse momentum is genuine, but a hadron or a photon from the final state may be falsely identified as a lepton. This background contribution is calculated using the DJANGO~\cite{DJANGO} program. 
\par
For the hadronic channel the production of multi--jet events in photoproduction and NC deep inelastic scattering are the most important Standard Model backgrounds.  The RAPGAP~\cite{RAPGAP} generator is used to model multi--jet production in NC deep inelastic scattering for virtualities of the exchanged photon $Q^2>4$~GeV$^2$ .
Multi--jet events with photon virtualities $Q^2<4$~GeV$^2$ are generated with the PYTHIA Monte Carlo program~\cite{PYTHIA}. Both generators rely on first order QCD matrix elements
and use leading--log parton showers
and string fragmentation~\cite{JETSET74}.
Both light and heavy quark flavours are generated. 
The GRV LO (GRV--G LO) parton densities~\cite{SFGRVGLO} in the proton
(photon) are used. The production of $W$ bosons and their hadronic decay are simulated using the EPVEC generator. This contribution is negligible in the present analysis.
\par
All generated events are passed through the full GEANT~\cite{Brun:1987ma} based simulation of the H1 apparatus and are reconstructed using the same program chain as for the data. 

%%%%%%%%%%%%%%%%%%%%%%%%%%%%%%%%%%%%%%%%%%%
\section{Experimental Conditions}
%%%%%%%%%%%%%%%%%%%%%%%%%%%%%%%%%%%%%%%%%%%
The analysis is based on $e^\pm p$ collisions recorded by the H1 experiment between 1994 and 2000. At HERA electrons or positrons with an energy $E_e$ of 27.6~\gev~collide with protons at an energy of 920 GeV, giving a centre--of--mass energy of~$\sqrt{s} = 319$~\gev. Up to 1997 the proton energy was 820~\gev, giving $\sqrt{s} = 301$~\gev.
The data correspond to an integrated luminosity of 37.0~pb$^{-1}$ in $e^+p$ scattering at $\sqrt{s} = 301$~\gev~($\mathcal{L}_{301} =$ 37.0~pb$^{-1}$),  together with 13.6~pb$^{-1}$ in $e^-p$ scattering and 67.7~pb$^{-1}$ in $e^+p$ scattering at $\sqrt{s} = 319$~\gev~($\mathcal{L}_{319} =$ 81.3~pb$^{-1}$).
\par
A detailed description of the H1 detector can be found in \cite{H1detector}.
Only those components  essential  for this analysis are briefly described 
here. The right handed Cartesian coordinate system used in the following
has its origin at the nominal primary $ep$ interaction vertex. The
proton direction defines the $z$ axis.  The polar
angle, $\theta$, and transverse momenta, $P_T$, are defined with respect to this axis. The region $\theta < 90^\circ$ is referred to as the ``forward'' region. 
The pseudorapidity is defined as $\eta = -\mathrm{ln}( \mathrm{tan} \frac{\theta}{2})$.
\par
The inner tracking system contains the central ($25^\circ < \theta < 155^\circ$) and forward ($7^\circ < \theta < 25^\circ$) drift chambers. It is used to measure the trajectories of charged particles and to determine the position of the interaction vertex. Particle transverse momenta are determined from the curvature of the trajectories in a solenoidal magnetic field of 1.15 Tesla.
\par
Hadronic final state particles as well as electrons and photons are absorbed in the highly segmented liquid argon calorimeter~\cite{h1cal} ($ 4^\circ < \theta < 154^\circ$), which is 5 to 8 hadronic interaction lengths deep depending on the polar angle. It includes an electromagnetic section which is 20 to 30 radiation lengths deep. 
Electromagnetic shower energies are measured with a precision of $\sigma (E) / E = 12\% / \sqrt{E/\mathrm{GeV}} \oplus 1\%$, hadronic shower energies with  $\sigma (E) / E = 50\% / \sqrt{E/\mathrm{GeV}} \oplus 2\%$, 
as determined in test beam measurements~\cite{h1testbeam}.
In the backward region ($153^\circ < \theta < 178^\circ$), the liquid argon calorimeter was complemented by a lead--scintillator backward electromagnetic calorimeter (BEMC) before 1995 and by a lead--scintillating fibre spaghetti calorimeter~(SPACAL)~\cite{h1spacal} since 1995.
\par
The calorimeter is contained within a superconducting coil and an iron return yoke, instrumented with streamer tubes, which is used as a muon detector and covers the range $ 4^\circ < \theta < 171^\circ$. Tracks of penetrating particles, such as muons, are reconstructed from their hit pattern in the streamer tubes and are detected with an efficiency of above 90\%.
The instrumented iron also serves as a backing calorimeter to measure the energies of hadrons that are not fully absorbed in the liquid argon calorimeter. 
\par
In the forward direction, muons are also detected in the forward muon system, a set of drift chambers covering the range  $3^\circ < \theta <17^\circ$. This detector measures the muon track momenta from their curvature in a magnetic field provided by a toroidal iron magnet.
\par
The trigger conditions for interactions leading to high transverse energy in the final state, as expected for top quark production, are mainly based on liquid argon calorimeter signals.
Events in the leptonic channel are triggered by their calorimetric missing transverse momentum. The trigger efficiency is 50\% (85\%) for events with a missing transverse momentum above 12~GeV (25~GeV). 
Events containing an electron with an energy of at least 10 GeV are triggered via the energy deposition in the electromagnetic calorimeter with an efficiency larger than 95\%. Events with muons may also be triggered by a set of triggers based on signals consistent with a minimum ionising particle in the muon system in coincidence with tracks in the inner tracking system. 
In the hadronic channel, the triggering of events with three or more high $P_T$ jets is based on the scalar sum of the transverse energy deposited in the liquid argon calorimeter. For events containing three jets with transverse momenta above 25 GeV, 20 GeV and 15 GeV respectively, the trigger efficiency is close to 100\%.

\section{Search for Single Top Production in the Leptonic Channel}
\label{sec:leptons}

The decay cascade $t\rightarrow b W \rightarrow b \ell \nu$ yields events with a lepton, missing momentum and a jet. The search for top quarks starts with the selection of events containing a high $P_T$ lepton and missing transverse momentum. In these events kinematic reconstruction of potential top quark decays is performed yielding a preselected data sample which serves as the basis for the final selection of top quark candidates. To discriminate single top production from Standard Model background processes, observables characteristic of top quark decays are used in a cut--based top selection and later in a multivariate likelihood analysis.   

\subsection{Events with isolated leptons and missing transverse momentum}
\label{sec:wpreselection}

The search for top quark decays in the leptonic channel is based on the selection of events with high $P_T$ leptons and missing transverse momentum described in~\cite{Andreev:2003pm}. 
Further details on this analysis can be found in~\cite{mireille,dave}. The selection yields a sample of candidates for leptonic $W$ boson decays. The following variables are used to characterise the events.

\begin{itemize}
\item $P_T^\ell$: the transverse momentum of the lepton.
Electron transverse momenta are calculated using calorimetric information together with vertex information from the trackers. Muon transverse momenta are measured from the curvature of the charged track detected in the  central tracker or in the forward muon detector.
 
\item $\theta^\ell$: the polar angle of the lepton. 

\item Charge of the lepton: The charge is measured from the track associated with the lepton. It is considered to be determined if the signed curvature of the track is different from zero with a measurement accuracy of better than two standard deviations. For less accurate measurements the lepton charge is labelled as ``undefined''.

\item $P_T^{miss}$: the total missing transverse momentum reconstructed from all observed final state particles.
% (electrons, muons and hadrons). 

\item $P_T^X$: the transverse momentum of the hadronic final state.
The hadronic final state, denoted by~$X$, is reconstructed by combining energy deposits in the calorimeter and charged tracks as described in~\cite{fscomb}. It does not include the energy deposited by any identified leptons in the event.

\end{itemize}

The main selection criteria are the requirement that there be an electron or muon with high transverse momentum $P_T^\ell >10$~GeV in the polar angle range $5^\circ < \theta^\ell<140^\circ$ and large missing transverse momentum $P_T^{miss}>12$~GeV as a signature of the undetected neutrino from the $W$ decay. 
The lepton is required to be isolated from neighbouring tracks or jets. The distance  $D_{track}$ in pseudorapidity-azimuth ($\eta$-$\phi$) space of the closest track from the lepton is required to be $>0.5$ and that of the nearest jet, $D_{jet}$, to be $>1.0$.
In the muon channel, an additional cut on the transverse momentum of the hadronic final state, $P_T^X>12$~GeV, is applied. Further selection criteria are applied to suppress processes where leptons are faked or missing transverse momentum is induced by measurement fluctuations, as discussed in detail in~\cite{Andreev:2003pm}. 
\par
In the full $e^\pm p$ data sample, $19$ events~\cite{Andreev:2003pm} are selected, compared with a Standard Model prediction of $14.5 \pm 2.0$, 
where the latter is dominated by $W$ production 
($10.7 \pm 1.8$).
One of the $19$ events was observed in $e^-p$ collisions. 

\subsection{\bf Kinematic reconstruction of the top quark decay}
\label{nureco}
In order to calculate the kinematics of a top quark decay, a reliable reconstruction of both the $b$ quark and the neutrino from the $W$ decay is necessary. 

%\subsubsection{$\mathbf b$--quark reconstruction}
\vspace*{0.5cm}
\noindent
{\bf Reconstruction of the \mbox{\boldmath$b$} quark }\\[0.5cm]
\noindent
In most cases the $b$ quark manifests itself as a single high $P_T$ jet in the detector. However, gluon radiation from the $b$ quark can lead to final states with more than one jet. Therefore the $b$ quark momentum is reconstructed as the sum of the momenta of  all jets found in the event. Jets are identified using an inclusive $k_T$ algorithm~\cite{Catani:1993hr} with a minimum jet transverse momentum of $4$~GeV. The sum of all jets gives a better approximation to the $b$ quark momentum than the full hadronic final state $X$, since it is less sensitive to particles originating from the proton remnant.

%\subsubsection{Neutrino reconstruction}
\vspace*{0.5cm}
\noindent
{\bf Reconstruction of the neutrino }\\[0.5cm]
\noindent
The transverse momentum vector of the neutrino corresponds to the vector of the total
missing transverse momentum: $\vec{P}_T^{\nu} = \vec{P}_T^{miss}$.
Concerning the longitudinal momentum of the neutrino, two cases are treated separately.
\begin{itemize}
\item {\it Tagged events:} \ the scattered beam electron
is detected in  one of the calorimeters.  This is expected to be the case for 30\% of single top events.
In the electron channel the scattered beam electron is assumed to be the one with the lower transverse momentum, which  according to the simulation is the correct choice in 95\% of top events. From energy and longitudinal momentum balance we obtain
\begin{equation}
  \left( E-P_z \right)^{\nu} = 2E_e - \left( E-P_z \right)^{leptons} - \left( E-P_z \right)^X \ .
\end{equation}
\item {\it Untagged events:} \ 
the scattered beam electron is lost in the beam pipe and
therefore its longitudinal momentum is unknown. In this case a constraint is applied on the invariant mass of the lepton and the neutrino from the $W$ decay:
\begin{equation}
\label{eq:mwconstraint}
 M_{\ell\nu} = \sqrt{P_\ell^2 + P_\nu^2 + 2P_\ell P_\nu}\ \approx \sqrt{2P_\ell P_\nu} = M_W = \ 80.42 \ \mbox{GeV} \ ~\cite{Hagiwara:fs} ,
\end{equation}
 where $P_\ell$ and $P_\nu$ denote the four--vectors of the lepton and the neutrino, respectively. The constraint on the $W$ mass generally yields two possible solutions for $\left( E-P_z \right)^{\nu}$. If two solutions for the neutrino kinematics exist, one solution corresponds to a backward neutrino with $\theta_\ell<\theta_\nu$, where $\theta_\nu$ is the neutrino polar angle, while the other solution corresponds to a forward neutrino with $\theta_\ell>\theta_\nu$. 
The solution which is most likely according to the model for anomalous top production is chosen. It is found that in top quark decays where the charged lepton is observed at small polar angles ($\theta_\ell<18^\circ$) the backward neutrino solution is favoured, while for large lepton polar angles ($\theta_\ell>40^\circ$) the forward neutrino solution is favoured~\cite{jochen}. 
In the intermediate region of the polar angle neither solution is favoured. Therefore, for $18^\circ<\theta_\ell<40^\circ$, the solution is selected that yields an invariant mass $M_{\ell\nu b}$ of the system consisting of the lepton, neutrino and $b$ candidate closest to the nominal top mass of $m_t=175$~GeV. 
\end{itemize}

 This neutrino reconstruction method is discussed in detail in~\cite{jochen}. It has a reconstruction efficiency of 99\% (95\%) for simulated top events in the electron (muon) channel. The widths from Gaussian fits of the top reconstructed mass $M_{\ell\nu b}$  distributions obtained for simulated top decays are $13~(18)$~GeV for the electron (muon) channel. 
\par
\begin{figure}[htb]
\centering
\epsfig{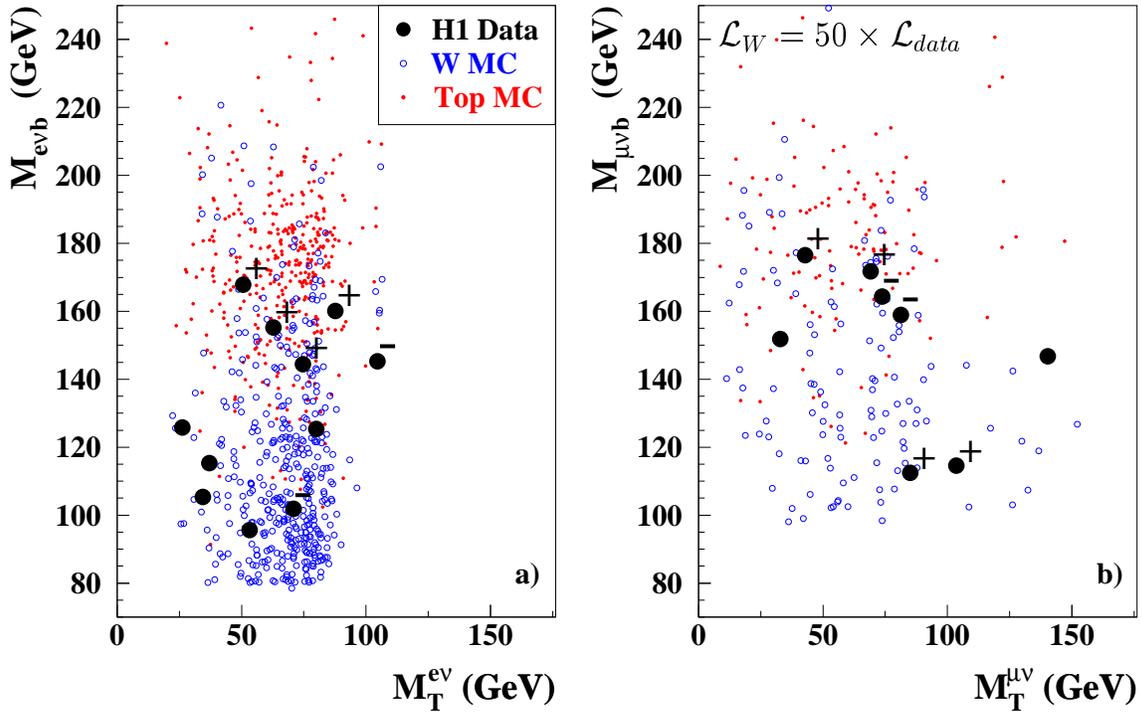} 
\caption{The invariant mass $M_{\ell\nu b}$ plotted against the lepton-neutrino transverse mass $M_T^{\ell\nu}$ for the isolated electron (a) and muon (b) events after the neutrino reconstruction. The data events (points) are compared with simulated top events produced via  FCNC interactions (small points, arbitrary normalisation) and events from Standard Model $W$ production (open circles) corresponding to 50 times the integrated luminosity of the data. 
For each data event the lepton charge is indicated, if it is determined with a measurement accuracy of better than two standard deviations.
}
\label{fig:mtop_mt}
\end{figure}

The kinematics of all $19$ isolated lepton events allow the reconstruction of a neutrino according to the above procedure. In figure~\ref{fig:mtop_mt} the invariant mass $M_{\ell\nu b}$ of these events is plotted against the lepton--neutrino transverse mass, defined as: 
$$M_T^{\ell\nu} = \sqrt{(P_T^\ell+P_T^\nu)^2-(\vec{P}_T^\ell+\vec{P}_T^\nu)^2} \ , $$
where $\vec{P}_T^\ell$ and $\vec{P}_T^\nu$ are the transverse momentum vectors of the lepton and the neutrino respectively. For each event the measured lepton charge is also indicated. Several events are situated at large masses $M_{\ell \nu b}$ close to the top mass and have transverse masses $M_T^{\ell\nu}$ compatible with there being a $W$ boson in the event. 

\subsection{Selection of top quark decays in the leptonic channel}
\noindent
{\bf Top preselection}\\[0.5cm]
\noindent
In addition to the kinematic reconstruction of the top quark decay, the lepton charge is also exploited. The decay chain \mbox{$t \rightarrow bW^+ \rightarrow b \ell^+ \nu_\ell$} produces only positively charged leptons. The production of anti--top quarks, which would yield negatively charged leptons, is strongly suppressed, as mentioned in section~\ref{theory}. To reduce the contribution from processes other than FCNC top quark production, negatively charged leptons are therefore rejected. 
The ``top preselection'' thus consists of the following three steps:
\begin{itemize}
\item selection of isolated lepton events with missing transverse momentum;
\item neutrino reconstruction;
\item rejection of leptons with negative charge.
\end{itemize}

The preselected top sample contains $9$ electron events and $6$ muon events compared with an expectation from Standard Model processes of $8.40 \pm 1.06$ and $1.88 \pm 0.32$, respectively. For negative lepton charges, $4$ events are found while $3.48 \pm 0.53$ are expected from Standard Model sources. The event yields, Standard Model predictions and selection efficiencies after each step of the top preselection are summarised in table~\ref{tableback}.
\par
 Candidates for single top event production are searched for in the preselected top sample. Both the cut--based top selection  described below and the multivariate likelihood analysis described in section~\ref{sec:multivar} exploit kinematic observables that are  characteristic for top quark decays.
% Candidates for single top event production are searched for in the preselected top sample, first using a cut--based selection described below. A multivariate likelihood analysis based on the same sample is also performed, as described in section~\ref{sec:multivar}. Both top selections exploit kinematic observables that are  characteristic for top quark decays.

%The observed and expected event yields for the three steps of the top preselection are summarised in table~\ref{tableback}.

%\newpage 

\vspace*{0.5cm}

\noindent
{\bf Observables for top quark decays in the leptonic channel}\\[0.5cm]
\noindent
\label{sec:cutbasedlep}
The following three observables are chosen to discriminate single top production from Standard Model $W$ production and other Standard Model processes.

\begin{enumerate}

\item $P_T^b$: the transverse momentum of the $b$ candidate. 
\item $M_{\ell \nu b}$: the invariant mass of the system consisting of the lepton, neutrino and $b$ candidate, which corresponds to the top quark candidate  mass. 
\item \thstarlep: the $W$ decay angle -- defined as the angle between the charged lepton momentum in the rest frame of the $W$ boson and the $W$ direction in the rest frame of the top quark.

\end{enumerate}

\begin{figure}[htb]
\centering 
\epsfig{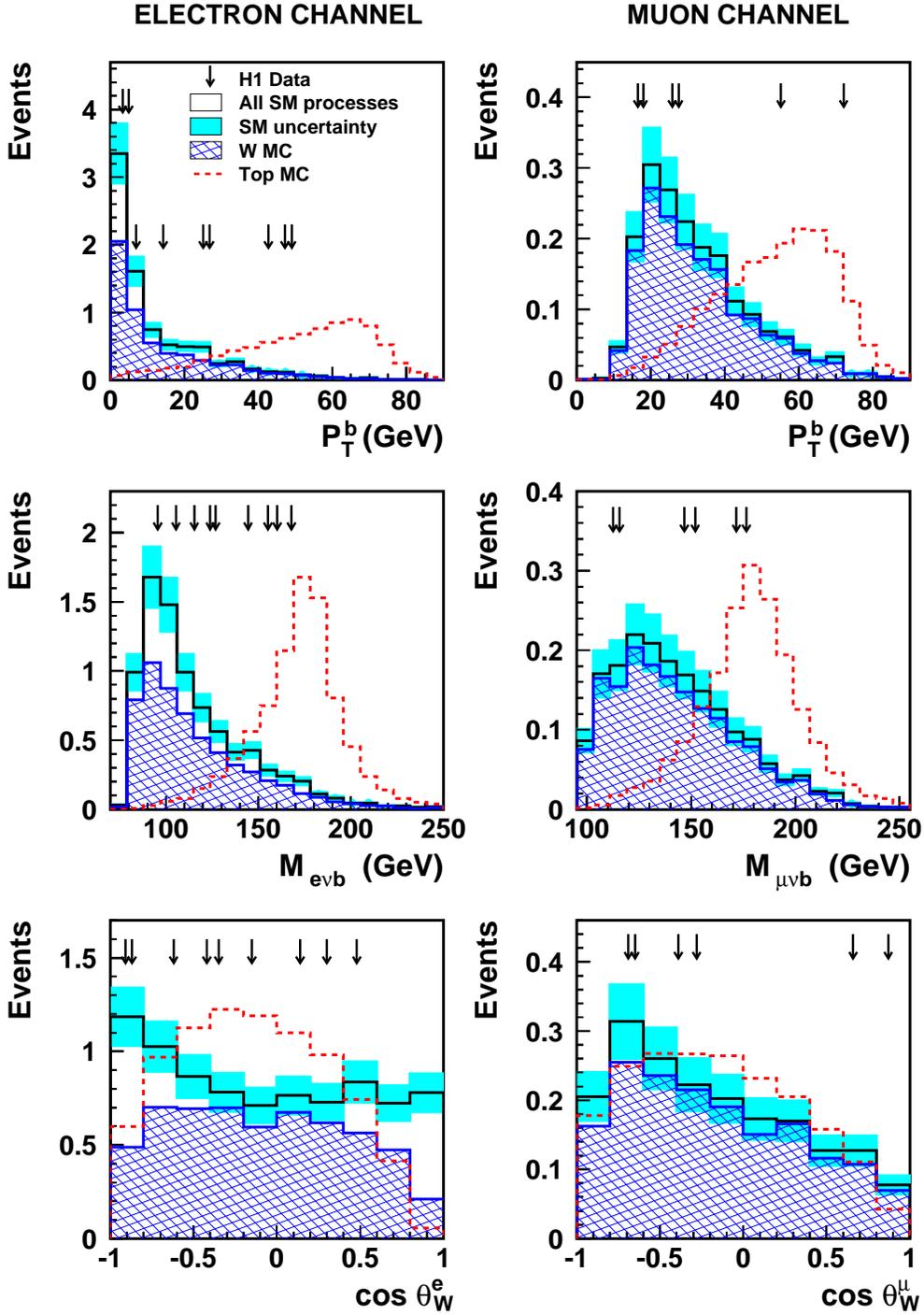}
\caption{
Distributions of the observables $P_T^b$, \mtopl~and \costarlep for the top preselection in the electron channel (left) and the muon channel (right). In each figure the arrows indicate the measured values for the data events: 9 events in the electron channel and 6 events in the muon channel. The solid histogram corresponds to the total Standard Model expectation. 
The hatched histogram represents the contribution from Standard Model $W$ production. The dashed histogram shows the distribution for simulated top events with an arbitrary normalisation.
}
\label{varl} 
\end{figure}

Distributions of the observables \ptb, \mtopl and \costarlep are shown in figure~\ref{varl} for the top preselection. Also shown are the signal distributions of simulated top events with an arbitrary normalisation. An excess of events at the highest values of $P_T^b$ is visible in both the electron and the muon channels. The masses  \mtopl for some of the electron and muon events are compatible with the top quark mass.  
\par

\vspace*{0.5cm}
\noindent
{\bf  Cut--based top selection in the leptonic channel }\\[0.5cm]
\noindent
Starting from the top preselected sample, the cuts that are used to select top candidates are \mbox{\ptb$>30$~GeV} and \mbox{\mtopl$>140$~GeV}. In this cut--based selection, no restriction is imposed on \costarlep,  since it does not yield an efficient separation of single top quark from Standard Model $W$ production, while the contribution from other Standard Model processes is already reduced to a negligible level by the \ptb and  \mtopl cuts. 
\par
In this cut--based analysis, $3$ electron events and $2$ muon events are selected as top quark candidates in the full $e^\pm p$ data sample. 
Some properties of these events are presented in table~\ref{tab_topmasses}. The  background expectation from Standard Model processes is  $0.65 \pm 0.10$ events for the electron channel and $0.66 \pm 0.12$ for the muon channel, as summarised in table~\ref{tableback}. The efficiency for simulated top events is $36\%$ ($38\%$) for the electron (muon) channel, taking into account the top decays where the $W$ boson decays via $W \rightarrow \tau \rightarrow e(\mu)$. Combining both channels, there are $5$ top quark candidates in the data for an expectation of $1.31$ $\pm$ $0.22$ from background processes.
\par
The systematic uncertainties on the Standard Model prediction that are relevant for the leptonic channel are described in~\cite{Andreev:2003pm}. They are dominated by the uncertainty of $15$\% on the NLO cross section calculation for Standard Model $W$ production~\cite{SPIRA}.

\section{Search for Single Top Production in the Hadronic Channel}
\label{sec:hadrons}
A search for single top production is also performed in the hadronic channel.
The decay cascade \mbox{$t\rightarrow b W \rightarrow b q\bar{q'}$} yields events with at least three jets with high transverse momenta. The main Standard Model background is the QCD production of high $P_T$ jets in photoproduction and neutral current deep inelastic scattering. 
First, a high statistics sample of multi--jet events is compared with simulations of Standard Model multi--jet production to check that the background is well understood. The search for top quarks is then performed in a multi--jet sample restricted to large transverse momenta (``top preselection''). Further details of the analysis can be found in~\cite{jochen}.

\subsection{Multi--jet events}

Since only LO Monte Carlo simulations with leading--log parton showers are used to model QCD multi--jet production, these simulations may give only  approximate descriptions of the shape and normalisation of kinematic distributions. Therefore, as a first step, the agreement in shape of the data and the simulations is studied using multi--jet events. 
The LO background simulations are normalized to the observed number of events in the data, to also account for the higher order QCD effects.
\par
Jets are reconstructed using the inclusive $k_T$ 
algorithm based on calorimetric energy deposits combined with well measured tracks.  To ensure a reliable measurement, only jets in the pseudorapidity range $-0.5<\eta^{jet}<2.5$ and with transverse momenta $P_T^{jet}>4$~GeV 
are considered. In order to remove electrons which are misidentified as jets, each jet is required to have either an electromagnetic energy fraction of less than 90\% or a jet size larger than 0.1, where the jet size is defined to be  the energy weighted average distance in the $\eta$--$\phi$ plane  of the particles composing the jet from the jet axis.
\par
\begin{figure}[htb]
\centering 
\epsfig{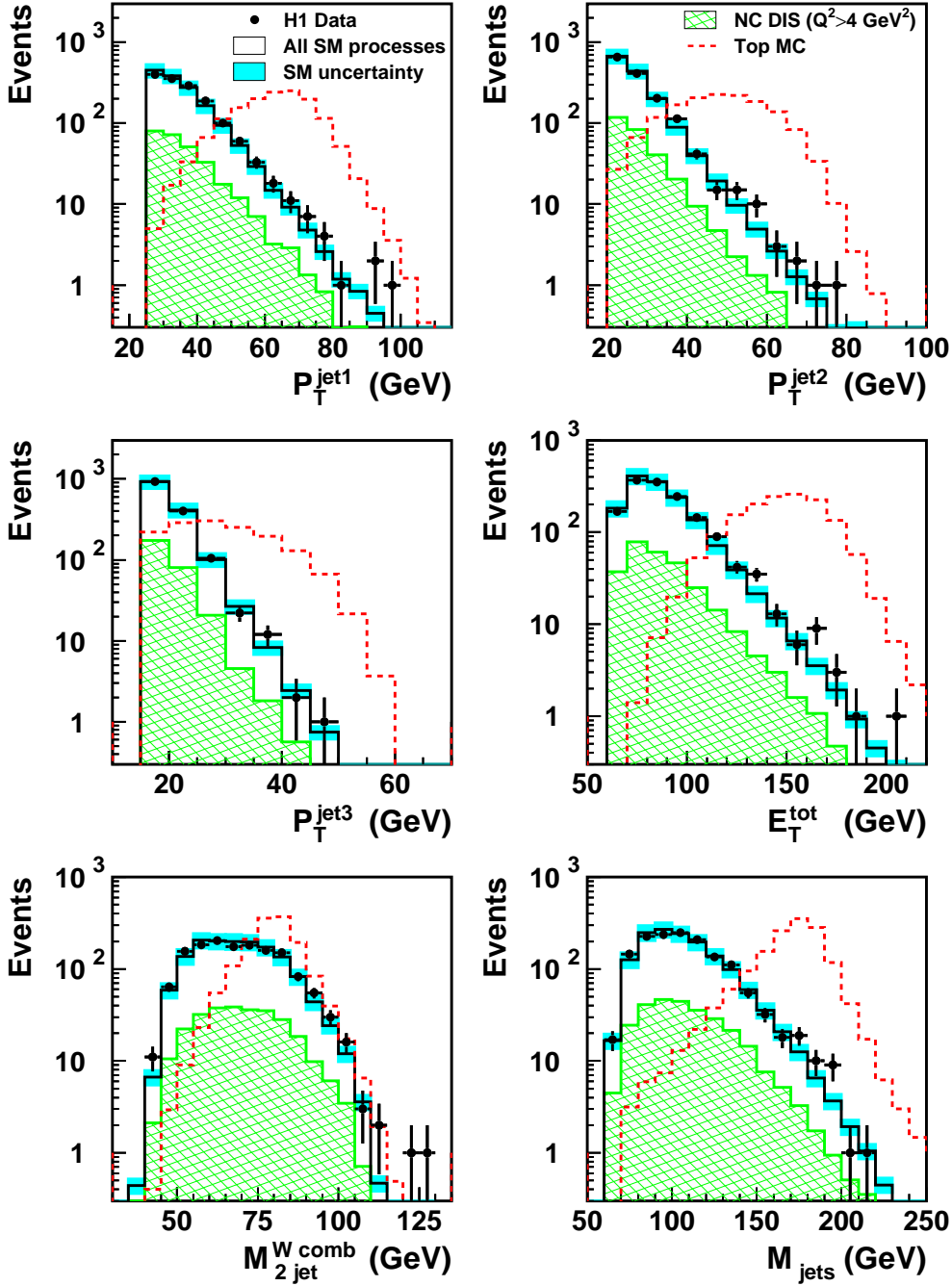}
\caption{Distributions of the multi--jet events with $P_T^{jet1} > 25$~GeV, $P_T^{jet2} >20$~GeV and $P_T^{jet3} >15$~GeV. The jet transverse momenta are shown for the three highest $P_T$ jets, as well as the total hadronic transverse energy, $E_T^{tot}$,  the invariant di--jet mass closest to the $W$ mass, $M_{2jet}^{W comb}$, and the invariant mass of all jets, $M_{jets}$. The data (symbols) are compared with the Standard Model simulations (histograms) after application of the normalisation factors of 1.29 (1.40) for PYTHIA (RAPGAP). The expectation from Standard Model processes is dominated by low $Q^2$ multi--jet production. The DIS contribution for $Q^2>4 \ \mathrm{GeV}^2$ is shown as a hatched histogram. The error band represents a systematic uncertainty of $18$\% on the total Standard Model prediction. The expected shape of the top signal is shown as the dashed histogram with an arbitrary normalisation.}
\label{jet:control} 
\end{figure}

 Events with at least three jets with $P_T^{jet1}>25$ GeV, $P_T^{jet2}>20$ GeV and $P_T^{jet3}>15$ GeV are used to study the agreement between the data and the simulations. The multi--jet data sample contains 1472 events.
In figure~\ref{jet:control}, distributions of various kinematic quantities are shown for these events and  compared with the Standard Model predictions and a simulation of single top production. The transverse momenta of the three highest $P_T$ jets, the total hadronic transverse energy ($E_T^{tot}$),
the di--jet invariant mass closest to the $W$ mass ($M_{2jet}^{W comb}$) and the invariant mass of all jets ($M_{jets}$) show good agreement with the Standard Model simulations in shape.
\par
The appropriate overall normalisation factors for the PYTHIA and RAPGAP simulations are determined using two complementary subsamples of the multi--jet events. One subsample contains events where no electron is identified (low $Q^2$ sample), where the prediction is dominated by the PYTHIA simulation. The other subsample contains events with an identified electron (high $Q^2$ sample), in which case the prediction is dominated by the RAPGAP simulation. A normalisation factor of~1.29 is applied to the event yield predicted by PYTHIA for $Q^2<4$~GeV$^2$ and a factor of~1.40 is applied to RAPGAP for $Q^2>4$~GeV$^2$. 
\par

\subsection{Selection of top quark decays in the hadronic channel}
\noindent
{\bf  Top preselection }\\[0.5cm]
\noindent
The search for top quarks in the hadronic channel is performed in a sample which is further restricted to the high transverse momentum region defined by  
$P_T^{jet1}>40$~GeV, $P_T^{jet2}>30$~GeV and $P_T^{jet3}>15$~GeV. Since top quarks typically deposit a large amount of transverse energy in the detector, a cut on the total hadronic transverse energy of $E_T^{tot}>110$~GeV is also applied. In addition, one of the jet pairings must yield an invariant mass between $65$~\gev~and $95$~\gev, corresponding to a window around the nominal $W$ mass with a width of twice the mass resolution obtained for hadronic $W$ decays.
These selection criteria are  referred to as 
the top preselection in the following. 
In the data $92$ events are selected. After application of the normalisation factors, the expectation for Standard Model processes is $92.4 \pm 16.6$. The good agreement between the data and the prediction in the top preselection indicates that the normalisation factors obtained for the multi--jet sample are also valid at high transverse momenta. 
\par
\begin{figure}[p]
  \begin{center}
     \epsfig{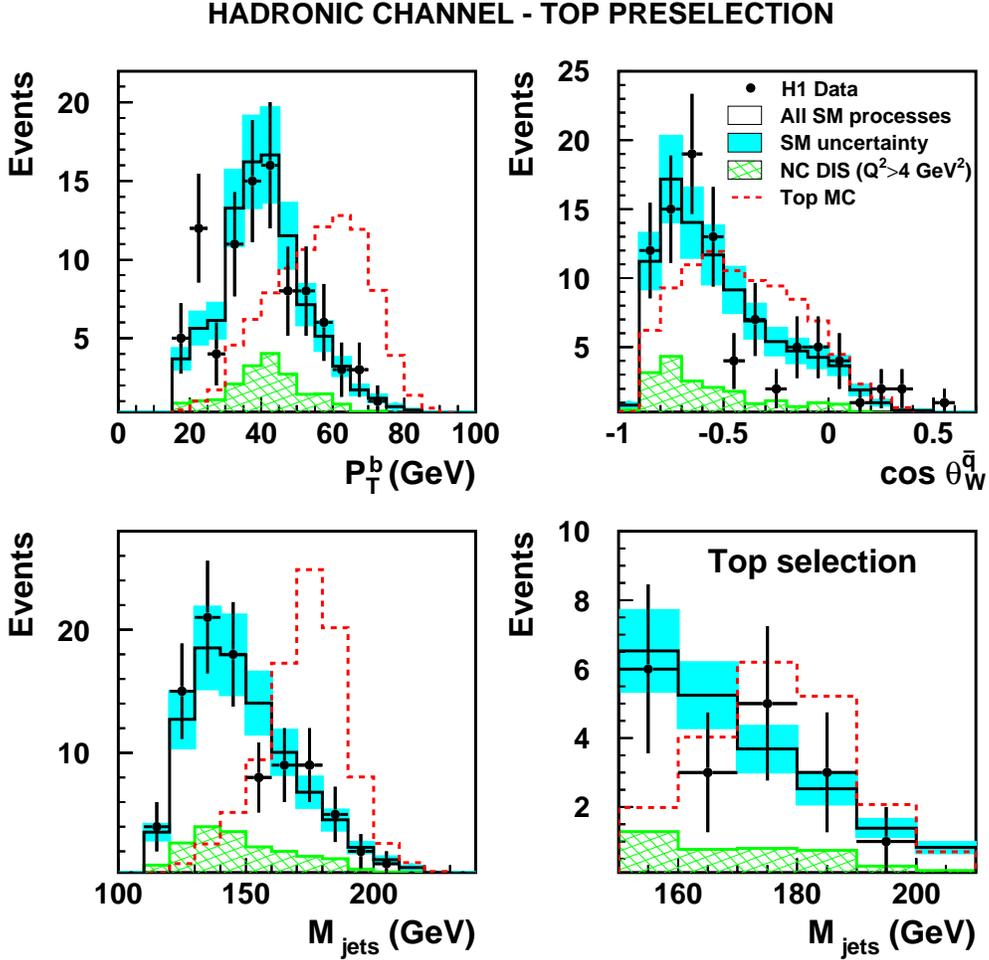}
  \end{center}
  \caption{Distributions of the observables  $P_T^{b}$, \costarhad and $M_{jets}$ for the top preselection in the hadronic channel. The data (symbols) are compared with the Standard Model simulations (histograms) after application of the normalisation factors of 1.29 (1.40) for PYTHIA (RAPGAP).  The expectation from Standard Model processes is dominated by low $Q^2$ multi--jet production. The DIS contribution for $Q^2>4 \ \mathrm{GeV}^2$ is shown as a hatched histogram. The error band represents the systematic uncertainty of $18$\% on the total Standard Model prediction. The expected shape of the top signal is shown as the dashed histogram with an arbitrary normalisation. The $M_{jets}$ distribution is also shown after the full cut--based top selection (lower right).}
  \label{jet:observables} 
\end{figure}

\vspace*{0.5cm}
\noindent
{\bf Observables for top quark decays in the hadronic channel}\\[0.5cm]
\noindent
\label{sec:cutbasedhad}
The observables used for the discrimination of the top signal from the QCD background are chosen in analogy to the leptonic channel (section~\ref{sec:cutbasedlep}). 
The jet among the three highest $P_T$ jets that is not used to form the $W$ mass is assigned to the $b$ quark (``$b$ candidate''). A study using simulated top events shows that this hypothesis correctly identifies the $b$ quark jet in 70\% of the events. The three characteristic observables used are: 
\begin{enumerate}
\item $P_T^{b}$: the transverse momentum of the $b$ candidate;
\item $M_{jets}$: the mass of the top quark -- reconstructed as the invariant mass of all jets in the event. The width of a Gaussian fit to the mass distribution obtained for simulated top decays is $14$~GeV; 
\item \thstarhad: the $W$ decay angle --  defined as the angle in the $W$ rest frame between the lower $P_T$ jet of the two jets associated to the $W$ decay and the $W$  direction in the top quark rest frame. The helicity structure of the decay implies that the lower $P_T$ jet corresponds to the $\bar{q}$ from the $W$ decay in most of the top events. 
\end{enumerate}
Distributions of these three observables are shown in figure~\ref{jet:observables} for  the top preselection and are compared with the Standard Model processes and with the simulated top signal. Good agreement between the data and the Standard Model simulations is seen for all three distributions. No sign of an excess compatible with single top production is visible. 

\vspace*{0.5cm}
\noindent
{\bf  Cut--based top selection in the hadronic channel }\\[0.5cm]
\noindent
In the hadronic channel harsher cuts  need to be applied than in the leptonic channel to enhance the top signal. 
To select top quark candidates following  the top preselection, 
the transverse momentum of the $b$ candidate has to fulfill  $P_T^{b}>40$~\gev~and the invariant mass of all jets has to be reconstructed in a  window around the top mass given by $150<M_{jets}<210$~GeV. In addition, \costarhad$>-0.75$ is required. 
For this selection the efficiency estimated with simulated top events is $30\%$. 
The number of candidate events selected is $18$,  compared with $20.2 \pm 3.6$ 
events expected from Standard Model processes. In figure~\ref{jet:observables} the reconstructed mass $M_{jets}$ of the selected top quark candidates is shown. The observed data events can be accounted for by Standard Model processes. 
\par
The main experimental systematic uncertainties on the expected number of events for the top selection are due to the uncertainty in the absolute hadronic energy calibration of the calorimeter ($\pm 4\%$) and the 
measurement of the polar angles of the jets ($\pm 20$~mrad), leading to a total experimental uncertainty of $11\%$. Systematic effects due to the uncertainty in the luminosity 
measurement and the trigger inefficiencies are negligible. 
The normalisation of the Standard Model simulation is taken from the data and its uncertainty of 10\% thus corresponds to the statistical uncertainty on the observed number of data events.  An additional uncertainty of 10\% accounts for differences between the shapes of the  kinematic distributions in the data and the simulations. 
All uncertainties presented above are added in quadrature. The total systematic uncertainty on the expected number of events in the hadronic channel amounts to $18\%$. 

%=======================
\section{Multivariate Likelihood Analysis}
%=======================
 \label{sec:multivar}
In addition to the cut--based top quark selections presented in sections~\ref{sec:cutbasedlep} and~\ref{sec:cutbasedhad}, a multivariate analysis is performed as an alternative approach to the search for single top production. 
The observables used to discriminate the top signal from the Standard Model background are combined to form a single discriminator based on the relative likelihood approach as defined in~\cite{opallike}. 
\par
In this framework a set of $i$ observables $V=\{V_i\}$ with the corresponding densities ${p^{ signal}_i}$ and ${p^{ background}_i}$, calculated from Monte Carlo samples for the signal and background respectively, are used for each event to calculate a discriminator:
$$
{\cal D}(V) =   \frac{ {\cal P}^{ signal}}{{\cal P}^{ signal}+ {\cal P}^{ background}}, \mbox{ where } {\cal P}={\cal C}(V) \prod_{i} p_i \ .
$$
Here ${\cal C}(V)$ denotes Gaussian correction factors used to correct for correlations between the variables $V_i$ as explained in detail in~\cite{opallike}. The discriminator ${\cal D}(V)$ is an approximation to the likelihood that an event is part of the signal rather than the background. 
\par
\begin{figure}[htb]
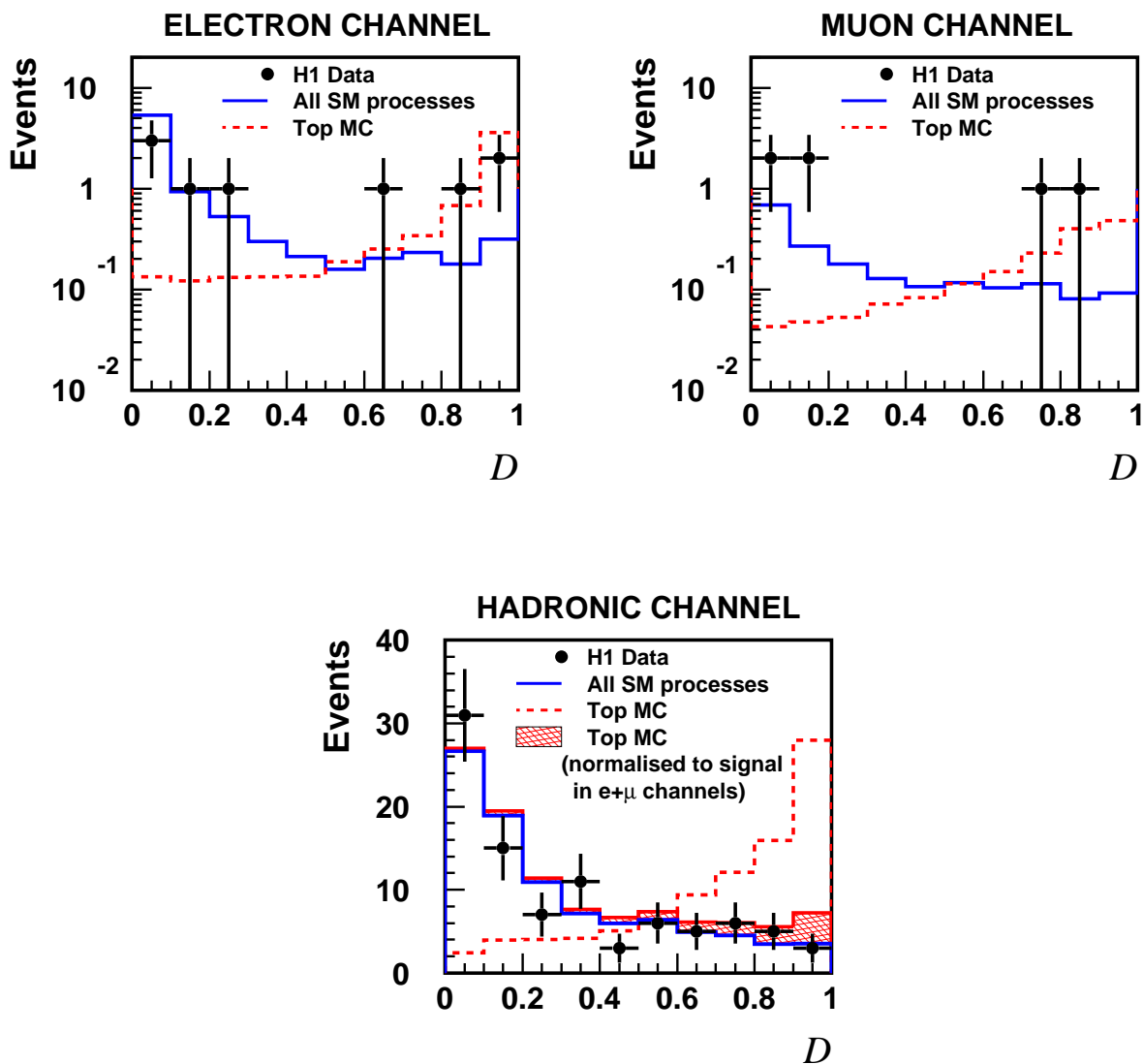

\centering 
  \begin{picture}(150,180)
    \put(-5,80){\epsfig{file=figure6a.eps,width=7cm}}
    \put(80,80){\epsfig{file=figure6b.eps,width=7cm}}
    \put(38,0){\epsfig{file=figure6c.eps,width=7cm}}
  \end{picture}\\[5mm]

\caption{
Distributions of the discriminator $\cal D$ for the data candidates (symbols) and the Standard Model expectation (histograms) for the top preselection in the electron, the muon and the hadronic channels. The solid histogram represents the Standard Model background. The dashed histogram shows the distribution for simulated top events with an arbitrary normalisation. For the hadronic channel, the top signal is also shown normalised according to the cross section derived in the combined electron and muon channels (hatched histogram) and added to the Standard Model histogram. 
}
\label{topfig_likl} 
\end{figure}

The variables used in this analysis are $V=\{ P_T^b, M_{\ell \nu b}, \cos{\theta_W^{\; \ell}}\}$ for the electron and muon channels and
$V=\{ P_T^{b}, M_{ jets}, \cos{\theta_W^{\; \bar{q}}}\}$ for the hadronic channel. 
The distributions of the discriminator variables for the signal and background processes according to the simulations and the distributions of the data events are shown in figure~\ref{topfig_likl} for the electron, muon and hadronic channels.
By definition, the top signal populates the region close to $\cal{D}$$\ =1$, while the Standard Model background populates the region close to $\cal{D}$$\ =0$. In both the electron and the muon channels, there are two populations in the data: one class of events that are more Standard Model--like and another class of events that are more top--like. The five top candidates selected in the cut--based analysis in section~\ref{sec:cutbasedlep} correspond to the events with the highest likelihood $\cal{D}$$\ > 0.7$. In the hadronic channel the distributions for the data events and for the QCD expectation are in good agreement. All candidates selected by the cut--based analysis in section~\ref{sec:cutbasedhad} are situated at $\cal{D}$$\ > 0.35$, but there is no evidence for any enhancement over the QCD background in this region. 
 A comparison between the likelihood discriminator analysis and the results of the cut--based analyses (sections~\ref{sec:cutbasedlep} and~\ref{sec:cutbasedhad}) for each channel is presented in table~\ref{res_cutlik} by applying a cut on the discriminator which yields the same signal efficiency as the cut--based selection.
\par
The probabilities for the Standard Model to fluctuate to discriminator distributions at least as unlikely as those observed in the data are evaluated using Monte Carlo experiments, following the approach proposed in~\cite{bockmethod}.
They are 0.3\% for the combined electron and muon channels, 40.1\% for the hadronic channel and 2.6\% for the combination of all channels.
\par
The  discriminator distributions are used to quantify the possible top signal contribution in the data using a maximum--likelihood fit.
A  likelihood function $L$ is introduced as the product of Poisson probabilities of observing $n_k$~data events in each bin~$k$ of the discriminator distribution: 
\begin{eqnarray}
\label{eq:fitlikelihood}
L = \prod_{k=1}^{n_{\mathrm{bin}}} e^{-\mu_k} \frac{\mu_k^{n_k}}{n_k!} \ ,
\end{eqnarray} 
where $\mu_k = B_k+S_k$ is the sum of the signal contribution~$S_k$ and expected background~$B_k$ in bin~$k$. The total top signal normalisation, $S = \sum_k S_k$, is fitted as a free parameter, while the normalisation of the background is fixed to the Standard Model prediction.  The value of $S$ which best matches the data can be obtained by maximising the likelihood function~$L$, or correspondingly by minimising the negative  log--likelihood function $-2 \ \mathrm{ln}L$. By using the factor $2$, the log--likelihood function corresponds to a $\chi^2$ function in the Gaussian limit. 
\par
The log--likelihood functions after subtraction of the minimum values 
are shown in figure~\ref{topfig_lfit} as functions of the single top cross section for the combined electron and muon channels, the hadronic channel and the combination of all channels. The conversion of the fitted signal normalisation~$S$ to a single top cross section at $\sqrt{s}=319$~GeV is done for each channel by folding in the corresponding  efficiency $\epsilon_{top}$, the top and $W$  branching ratio product $\mathcal{B}_{t \rightarrow b W} \cdot \mathcal{B}_{W\rightarrow ff'}$ and taking into account the  integrated luminosities 
$\mathcal{L}_{301}$ and $\mathcal{L}_{319}$:
\begin{equation}
\sigma(\sqrt{s} = 319~\gev) = \frac{S}{\epsilon_{top} \cdot \mathcal{B}_{t \rightarrow b W} \cdot \mathcal{B}_{W\rightarrow ff'} } \cdot \frac{1}{0.70 \cdot \mathcal{L}_{301}+\mathcal{L}_{319}}.
\end{equation}
Here, the factor $0.70$ is the ratio of the cross sections at $\sqrt{s}=301$~GeV and  $319$~GeV~\cite{Belyaev:2001hf}.
The branching ratio for $t \rightarrow b W$ is assumed to be $\mathcal{B}_{t \rightarrow b W}=100 \%$, 
in accordance with~\cite{Abe:1997fz}. 
 The combination of the different channels is performed by adding the log--likelihood functions of the single channels. 
In order to propagate the systematic uncertainties related to the measurement through the signal fitting procedure,
each observable affected by a systematic uncertainty is smeared according to a Gaussian distribution with a width corresponding to the size of the uncertainty. The full analysis is then repeated for a large number of values of each smeared observable. The r.m.s of the resulting distribution of shifts in the fitted cross section is taken as the corresponding systematic uncertainty. The uncertainties on the cross sections are added in quadrature for each error source. In order to include these uncertainties in the log--likelihood function, the function is approximated by a half--parabola on each side of the minimum (Gaussian approximation) and the width of the parabola is increased according to the total systematic uncertainty. 
As can be seen in figure~\ref{topfig_lfit}, the impact of the systematic uncertainties is most important for the hadronic channel, while it is negligible compared with the statistical uncertainties in the leptonic channels.

%=======================
\section{Cross Sections and Limits}
%=======================
\label{sec:results}

\begin{figure}[htb]
\centering 
\epsfig{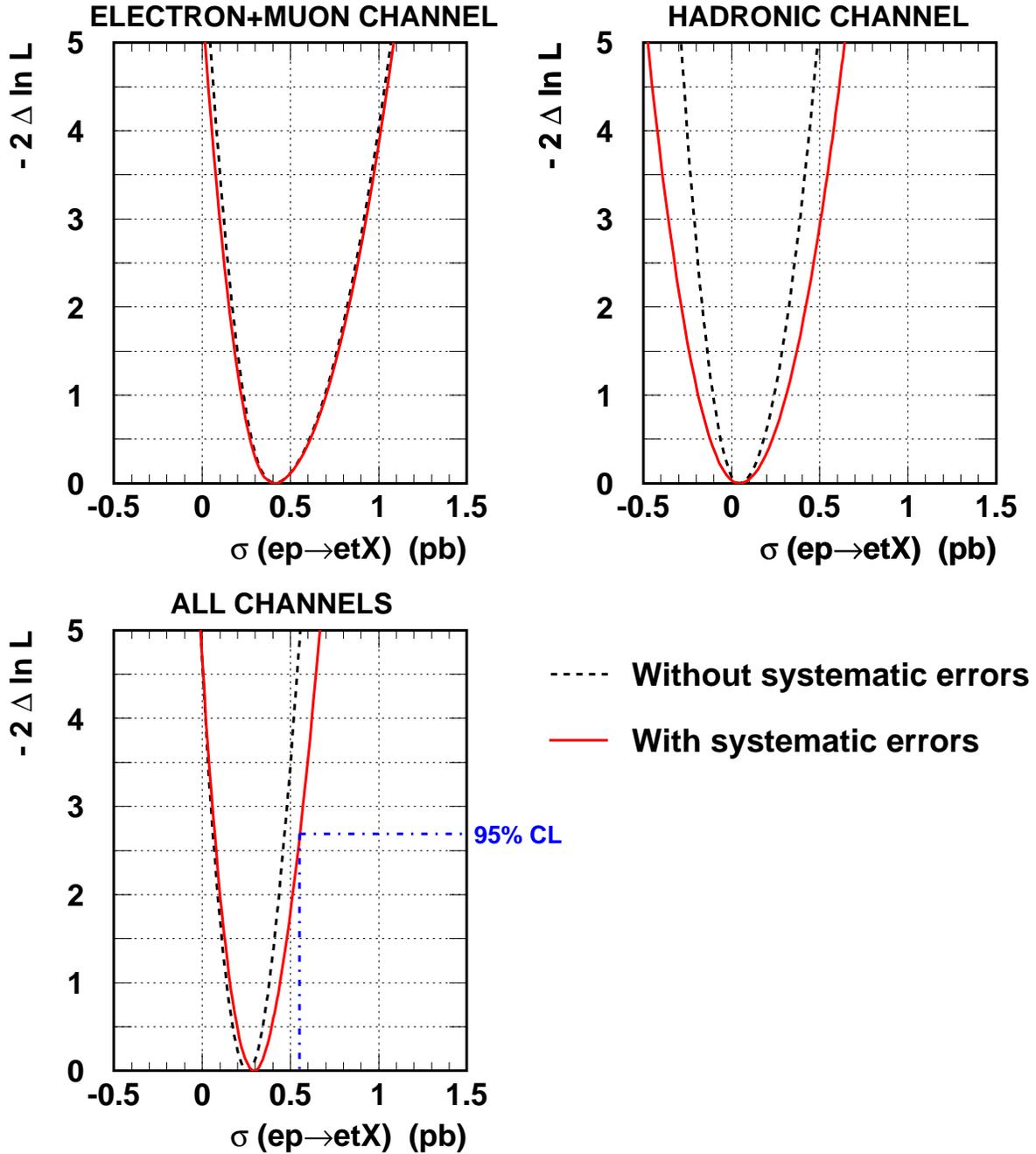}
\caption{
The zero--suppressed log--likelihood function $-2\Delta\mathrm{ln}L$ as functions of the single top cross section at $\sqrt{s}=319$~GeV for the combined electron and muon channels, the hadronic channel and the combination of all channels. For the latter, the one--sided (upper) exclusion limit at the 95\% confidence level on the single top cross section is marked by the dashed--dotted line.
}

\label{topfig_lfit} 
\end{figure}

The log--likelihood function for the combination of the electron and muon channels in figure~\ref{topfig_lfit} yields a single top cross section of $0.41^{+0.29}_{-0.19}$~pb at $\sqrt{s} = 319$~\gev. This cross section is different from zero by more than two standard deviations and reflects the observation of the five top quark candidate events presented in section~\ref{sec:leptons}. 
For the hadronic channel the log--likelihood function in figure~\ref{topfig_lfit} yields a single top cross section of $0.04^{+0.27}_{-0.23}$~pb at $\sqrt{s} = 319$~\gev. This cross section is consistent with no top signal, in accordance with the results of the cut--based analysis reported in section~\ref{sec:cutbasedhad}. Taking into account the statistical and systematic uncertainties, the results from the hadronic channel and the combined electron and muon channels are compatible at the $1.1~\sigma$ level.
The combination of all three channels yields a single top cross section of $0.29^{+0.15}_{-0.14}$~pb at $\sqrt{s} = 319$~\gev. The addition of a contribution from anomalous single top production to the Standard Model background provides a good description of the data. 
As discussed in section~\ref{theory}, the cross section for anomalous top production at HERA is approximately proportional to the anomalous magnetic coupling squared, $\sigma(e p \rightarrow e + t + X) \propto \kappa_{tu\gamma}^2$. The value obtained for  $\kappa_{tu\gamma}$ is $0.20^{+0.05}_{-0.06}$.
\par
In view of the small number of top candidates an upper limit on the single top production cross section is  calculated. 
It can be directly obtained from the log--likelihood functions presented in figure~\ref{topfig_lfit}. A one--sided exclusion limit at the 95\% confidence level corresponds to an increase of 2.69 units in $-2\Delta\mathrm{ln}L$. The resulting upper bound on the single top cross section at $\sqrt{s} = 319$~GeV for the combination of all analysed channels is
$$ \sigma(e p \rightarrow e + t + X, \sqrt{s} = 319~\GeV)
   < 0.55 \;{\mbox{pb}} \; ( 95 \% {\mbox{ CL}}).$$ 
The bound on the top cross section is translated into an upper limit on the anomalous $tu \gamma$ coupling of:
$$ | \kappa_{tu\gamma} | < 0.27 \; (95 \% {\mbox{ CL}}) . $$ The limits obtained separately for the combined electron and muon channels and for the hadronic channel only are given in table~\ref{tab:limits}.
As a cross check, other statistical methods  have also been used to derive the exclusion limits, for example a Bayesian approach with a flat prior~\cite{Hagiwara:fs} or likelihood--based 
approaches~\cite{bockmethod,Junk:1999kv}. The results are consistent within $15\%$. 
\par
\begin{figure}[htb]
\centering
\epsfig{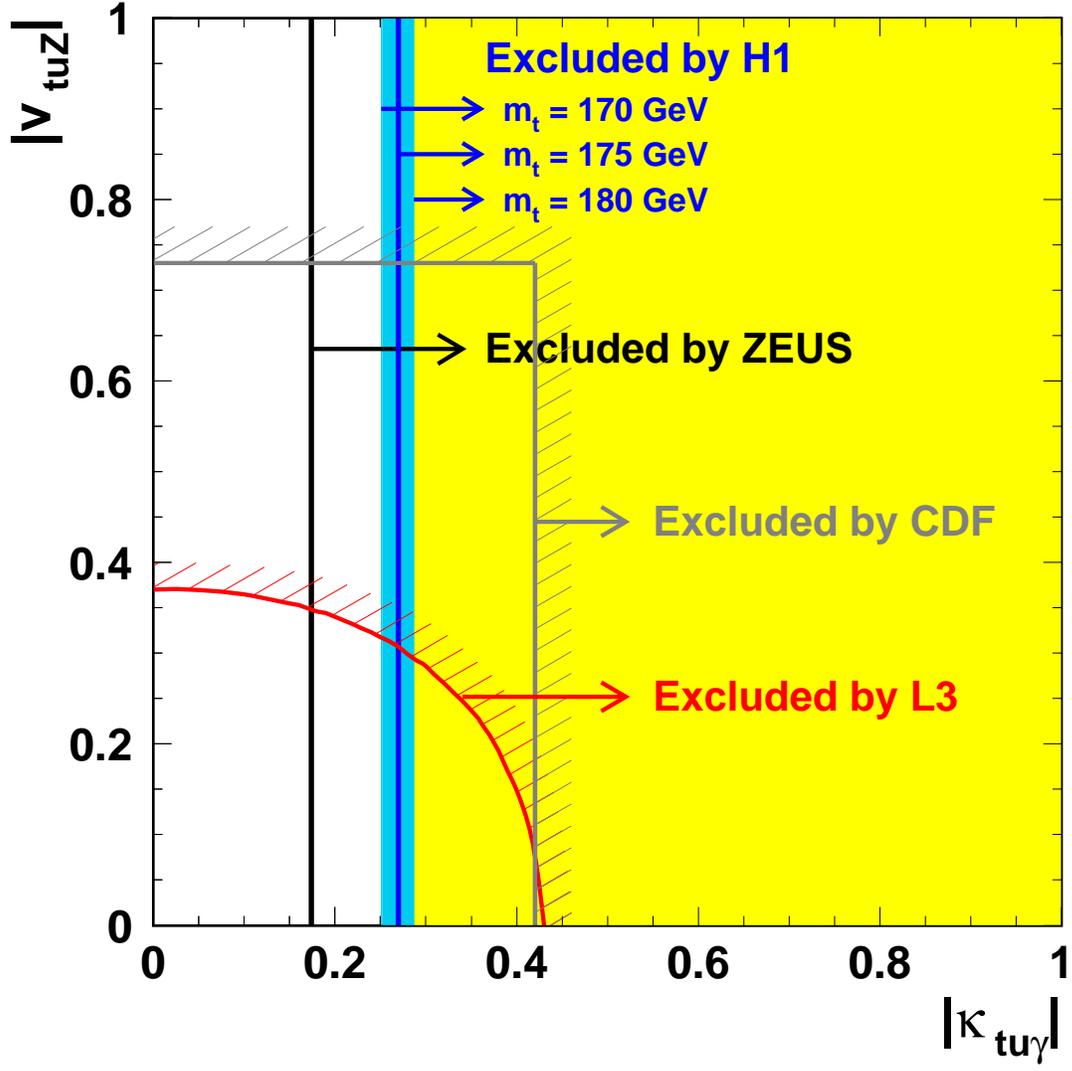} 
\caption{Exclusion limits at the 95\% confidence level on the anomalous $tq\gamma$ magnetic coupling $\kappa_{tu\gamma}$ and the vector coupling $v_{tuZ}$ obtained 
at the TeVatron (CDF experiment~\cite{Abe:1997fz}), 
LEP (L3 experiment is shown, which currently gives the best limit of the LEP experiments~\cite{LEP})
 and HERA (H1 and ZEUS experiments). 
The anomalous couplings to the charm quark are neglected $\kappa_{tc\gamma}=v_{tcZ}=0$.
%The H1 and ZEUS limits apply to the coupling $\kappa_{tu\gamma}$ to the $u$--quark only. 
The error band on the H1 limit shows the uncertainty on the coupling $\kappa_{tu\gamma}$ induced by a variation of the nominal top quark mass by $\pm 5$~GeV.
}
\label{fig:limits}
\end{figure}

The present limit on $\kappa_{tu\gamma}$  is consistent with the result obtained by the ZEUS collaboration~\cite{zeus_singletop} in the framework of the NLO QCD calculation~\cite{Belyaev:2001hf} for anomalous single top production: $ | \kappa_{tu\gamma} | < 0.17$ (95\% CL).
The limits on the top quark anomalous couplings obtained by the HERA experiments can be compared with the limits obtained from the search for single top production at LEP~\cite{LEP} and from the analysis of radiative top decays by the CDF collaboration~\cite{Abe:1997fz}.
Figure~\ref{fig:limits} represents the current status of the constraints on $\kappa_{tu\gamma}$ and $v_{tuZ}$. The limit on the anomalous coupling $\kappa_{tu\gamma}$ obtained in the present analysis significantly improves the CDF and LEP upper bounds if  the vector coupling $v_{tuZ}$ is not too large.
 The error band on the H1 limit represents the uncertainty induced by a variation of the nominal top quark mass of $m_t=175\gev$ by $\pm 5$~GeV in the analysis. Other theoretical errors are neglected. The H1 results on single top production are not in contradiction with the limits set by other experiments. 

\section{Summary}

A search for single top production is performed using the data sample collected by the H1 experiment between 1994 and 2000, corresponding to a total luminosity of 118.3~pb$^{-1}$. This search is motivated by the previous observation of 
events containing an isolated lepton, missing transverse momentum and large hadronic transverse momentum, a topology typical of the semileptonic decay of the top quark.
 \par
In a cut--based analysis,
$5$  events are selected as top quark candidate decays in the leptonic channel.
The prediction for Standard Model processes is $1.31$ $\pm$ $0.22$ events. 
The analysis of multi--jet production at high $P_T$, corresponding to a search for single 
top production in the hadronic channel, shows good agreement with the expectation for Standard Model processes within the uncertainties. 
\par
In order to extract the top quark production cross section, a multivariate likelihood analysis is performed in addition to the cut--based analyses. The top signal contribution in each channel is determined in a maximum--likelihood fit to the likelihood discriminator distributions. 
The results from the hadronic channel do not rule out a single top interpretation of the candidates observed in the electron and muon channels. For the combination of the electron, muon and hadronic channels a cross section for single top production of $\sigma=0.29^{+0.15}_{-0.14}$~pb at $\sqrt{s} = 319$~\gev~is obtained. This result is not in contradiction with limits obtained by other experiments. 
The addition of a contribution from a model of anomalous single top production yields a better description of the data than is obtained with the Standard Model alone. 
\par
%Due to the small number of top candidates, exclusion limits for the single top cross section of $\sigma<$~0.55~pb at $\sqrt{s}=319$~\gev~and for the anomalous $tu\gamma$ coupling of $| \kappa_{tu\gamma} | < 0.27$ are also derived at the 95 \%  confidence level. The HERA bounds extend into a region of parameter space so far not covered by  experiments at  LEP and the TeVatron.
Assuming that the small number of top candidates are the result of a 
statistical fluctuation, exclusion limits for the single top cross section of $\sigma<$~0.55~pb at $\sqrt{s}=319$~\gev~and for the anomalous $tu\gamma$ coupling of $| \kappa_{tu\gamma} | < 0.27$ are also derived at the 95 \%  confidence level. The HERA bounds extend into a region of parameter space so far not covered by  experiments at  LEP and the TeVatron.

\section*{Acknowledgements}

We are grateful to the HERA machine group whose outstanding
efforts have made this experiment possible. 
We thank
the engineers and technicians for their work in constructing and
maintaining the H1 detector, our funding agencies for 
financial support, the
DESY technical staff for continual assistance, 
and the DESY directorate for support and for the
hospitality which they extend to the non--DESY 
members of the collaboration. 
We thank A.~S.~Belyaev, P.~Bock, K.--P.~Diener and M.~Spira for useful discussions and fruitful collaboration.

%============================

\vspace*{3cm}

%%%%%%%%%%%%%%%%%%%%%%%%%%%%%%%%%%%%%%%%%%%%%%%%%%%%%%%%%%%%%%%%%%%%
%%%%%%% TABLES
%%%%%%%%%%%%%%%%%%%%%%%%%%%%%%%%%%%%%%%%%%%%%%%%%%%%%%%%%%%%%%%%%%%%%
%\input{rateallnew}
\begin{table}[hhh]
\begin{center}
\renewcommand{\arraystretch}{1.55}
\begin{tabular}{|cc|l|c|c|c|c|c|} \hline

\multicolumn{2}{|c}{} &\multicolumn{1}{l|}{\bf Electron Channel}& Data & Standard Model & $W$ only       & Top efficiency \\ \hline
&& Isolated Lepton + $P_T^{miss}$ & 11   & $11.53\pm 1.49$ & $8.17\pm 1.35$ & 47\% \\ \cline{3-7}
&&$\nu$ reconstruction & 11    & $11.06\pm 1.43$ & $7.93\pm 1.31$ & 46\%  \\ \cline{3-7}
\rule{0.2cm}{0pt}
\begin{rotate}{90} Top 
\end{rotate} 
&
\begin{rotate}{90} preselection
\end{rotate} 
\rule{0.2cm}{0pt}
&Cut on lepton charge & 9    & $8.40\pm 1.06$  & $5.72\pm 0.95$ & 45\%  \\ \hline
\multicolumn{2}{|c}{} &\multicolumn{1}{l|}{Top  cut--based selection}       & 3    & $0.65\pm 0.10$  & $0.57\pm 0.10$ & 36\%  \\ 
\hline
\hline
\multicolumn{2}{|c}{}&\multicolumn{1}{l|}{\bf Muon Channel}   & Data & Standard Model & $W$ only       & Top efficiency \\ \hline
&&Isolated Lepton + $P_T^{miss}$ &  8   & $2.96\pm 0.50$ & $2.54\pm 0.49$ & 46\% \\ \cline{3-7}
&&$\nu$ reconstruction &  8   & $2.70\pm 0.46$ & $2.38\pm 0.46$ & 44\%  \\ \cline{3-7}
\rule{0.2cm}{0pt}
\begin{rotate}{90} Top 
\end{rotate} 
&
\begin{rotate}{90} preselection
\end{rotate} 
\rule{0.2cm}{0pt}
&Cut on lepton charge &  6   & $1.88\pm 0.32$ & $1.67\pm 0.32$ & 43\%  \\ \hline
\multicolumn{2}{|c}{} &\multicolumn{1}{l|}{Top cut--based selection}        &  2   & $0.66\pm 0.12$ & $0.59\pm 0.12$ & 38\%  \\ 
\hline

\end{tabular}
\caption{Observed and predicted numbers of events for the three steps in 
the top preselection and for the cut--based top selection in the leptonic channel.  The ``W only'' column gives the prediction from Standard Model $W$ production alone. The numbers are presented for the full $e^\pm p$ data sample corresponding to an integrated luminosity of $118.3$ pb$^{-1}$.}
\label{tableback}
\end{center}
\end{table}

\begin{table}[htb]
\begin{center}
\renewcommand{\arraystretch}{1.3}
\begin{tabular}{|llc|l|c|c|c|} \hline
 Run & Event    & Lepton  & Charge       & $P_T^b$      & \mtopl                    & $M_T^{\ell\nu}$  \\
     &          & type    &              &              & $M_W$-constraint solution &   \\
     &          &         &              &(GeV)         & (GeV)                     & (GeV)    \\ \hline
248207 &  32134 & $e$    & + ($15\sigma$)  & 43    & $155^{+7}_{-7}$                  & $63^{+2}_{-2}$       \\ % 248207 32134
252020&  30485 & $e$    & + ($40\sigma$) & 47    & $168^{+8}_{-8}$                    & $51^{+2}_{-2}$       \\ % 252020 30485 tagged
268338 & 70014 & $e$    & + ($5.1\sigma$)  & 48    & $160^{+6}_{-6}$                  & $88^{+2}_{-2}$       \\ % 268338 70014
\hline
186729 &  702 & $\mu$   & + ($4.0\sigma$)  & 72    & $176^{+9}_{-12}$                  & $43^{+13}_{-22}$  \\ % 186729 702
266336&  4126 & $\mu$   & + ($26\sigma$) & 55     & $172^{+9}_{-10}$                   & $69^{+2}_{-3}$       \\ % 266336 4126
\hline
\end{tabular}
\caption{
Kinematics and lepton charges of the five top quark candidates in the leptonic channel. In one event (Run 252020 Event 30485) the scattered electron is detected and the ``tagged'' mass solution can be obtained for the top quark mass  $168^{+11}_{-11}$~GeV. The mass of the lepton-neutrino system of this event is measured to be  $M_{\ell\nu} = 79^{+12}_{-12}$~GeV.
}
\label{tab_topmasses}
\end{center}
\end{table}

\begin{table}[ht]
\begin{center}
\renewcommand{\arraystretch}{1.3}
\begin{tabular}{|c|c|c|c|c|c|c|}
\hline
                  & \multicolumn{2}{|c|}{Cut--based analysis}  &  \multicolumn{2}{|c|}{${\cal D} > {\cal D}_{min}$ }& ${\cal D}_{min}$ & Efficiency \\ \cline{2-5}
                  & Data          &  SM             & Data          &  SM              &          &         \\ \hline
Electron Channel  & $3$           & $0.65 \pm 0.10$ & $3$           &  $0.67 \pm 0.13$ & $0.72$     & $36 \%$ \\ \hline 
Muon Channel      & $2$           & $0.66 \pm 0.12$ & $2$           &  $0.62 \pm 0.12$ & $0.40$     & $38 \%$ \\ \hline 
Hadronic Channel  & $18$          & $20.2 \pm 3.6$  & $20$          &  $17.5 \pm 3.2 $ & $0.58$   & $30 \%$ \\ \hline 
%Hadronic Channel  & $18$          & $20.2 \pm 3.6$  & $18$          &  $18.0 \pm 3.2 $ & $0.60$   & $30 \%$ \\ \hline 

\end{tabular}
\end{center}
  \caption{Observed and predicted numbers of events in the cut--based top selection, compared with the selection using the single cut on the likelihood discriminator (${\cal D} > {\cal D}_{min}$) which yields the same efficiency for the top signal. The numbers are presented for the full $e^\pm p$ data sample corresponding to an integrated luminosity of $118.3$ pb$^{-1}$.}
  \label{res_cutlik}
\end{table}

\begin{table}[htb]
\begin{center}
\renewcommand{\arraystretch}{1.3}
\begin{tabular}{|c|c|c|} \hline
                 & $\sigma$  & $| \kappa_{tu\gamma} |$  \\ \hline 
Electron+Muon Channel &     $< 0.90$ pb           &  $< 0.35$                      \\ \hline
Hadronic Channel      &     $< 0.48$ pb           &  $< 0.25$                      \\ \hline
All Channels          &     $< 0.55$ pb           &  $< 0.27$                      \\ \hline

\end{tabular}
\caption{Exclusion limits at the 95\% confidence level  for the single top cross--section at $\sqrt{s} = 319$~GeV and for the anomalous $tu\gamma$ coupling.}
\label{tab:limits}
\end{center}
\end{table}

%%%%%%%%%%%%%%%%%%%%%%%%%%%%%%%%%%%%%%%%%%%%%%%%%%%%%%%%%%%%%%%
%%%%%%% figures
%%%%%%%%%%%%%%%%%%%%%%%%%%%%%%%%%%%%%%%%%%%%%%%%%%%%%%%%%

%%%%%%%%%%%%%%%%%%%%%%%%%%%%%%%%%%%%%%%%%%%%%%%%%%%%%%%%%%

\end{document}